# Artificial Neural Network to predict mean monthly total ozone in Arosa, Switzerland

Surajit Chattopadhyay* and Goutami Bandyopadhyay
1/19 Dover Place
Kolkata-700 019
West Bengal
India
*E-mail surajit_2008@yahoo.co.in

## Abstract

Present study deals with the mean monthly total ozone time series over Arosa, Switzerland. The study period is 1932-1971. First of all, the total ozone time series has been identified as a complex system and then Artificial Neural Networks models in the form of Multilayer Perceptron with back propagation learning have been developed. The models are Single-hidden-layer and Two-hidden-layer Perceptrons with sigmoid activation function. After sequential learning with learning rate 0.9 the peak total ozone period (February-May) concentrations of mean monthly total ozone have been predicted by the two neural net models. After training and validation, both of the models are found skillful. But, Two-hidden-layer Perceptron is found to be more adroit in predicting the mean monthly total ozone concentrations over the aforesaid period.

## 1. Introduction

The photochemical processes leading to formation of ozone ($O_3$) are highly complex in nature. Ozone is a secondary pollutant and is not usually emitted directly from stacks, but instead is formed in the atmosphere as a result of reactions between other pollutants emitted mostly by industries and automobiles. The ozone precursors are generally divided into two groups, namely oxides of nitrogen ($NO_X$) and volatile organic components (VOC) like evaporative solvents and other hydrocarbons. In suitable ambient meteorological condition (e.g. warm, sunny/clear day) ultraviolet radiation (UV) causes the precursors to interact photochemically in a set of reactions that result in the formation of ozone [3,4]. The process of ozone formation can be expressed as [19]:

$NO_2 + UV \rightarrow NO + O$

$O + O_2 + M \rightarrow O_3 + M$

Where, M is a third body molecule that remains unchanged in the reaction.

Ozone produced this way gets simultaneously destroyed as:

$O_3 + D \rightarrow DO + O_2$

Where, D implies additional reactant that destroys the ozone via oxidation.

Ozone absorbs both incoming solar radiation in the UV and visible region, and terrestrially emitted infrared (IR) radiation. Stratospheric ozone absorbs about $12 Wm^{-2}$ of solar radiation and $8 Wm^{-2}$ of terrestrial IR radiation. Almost 60% of this absorbed IR is radiated. Because of its capability to absorb the incoming radiation, the stratospheric ozone is a major source of stratospheric heating, which further heats the troposphere. Again, because of radiation of IR the tropopause gets some cooling. Thus, stratospheric ozone exerts both heating and cooling effect on the land-troposphere system. For ozone in the troposphere, however, both direct solar absorption and IR trapping warm the surface-troposphere system.

Because of its dependency upon weather conditions and precursor emissions formation of ozone is a highly complex and non-linear phenomenon. Various authors already established chaotic behavior of weather variables. The precursor emissions are also associated with dispersion conditions like wind speed, atmospheric stability etc. Furthermore, rates of reaction among pollutants vary significantly with meteorological conditions and availability of UV radiation. Thus, formation of ozone is immensely complex and non-linear phenomena.

From the previous discussion, it can be said that ozone, which acts as a shield to UV in the stratosphere, is good for stratosphere but hazardous for troposphere (0-10 Km high). Ozone is a lung irritant and a phototoxycant. It is responsible for crop damage, and is suspected of being a contributor to forest decline in Europe and parts of United States [19]. Studies revealed that since the end of 50s tropospheric ozone increased in the northern hemisphere [8]. This increase is contributed by intrusion of ozone rich stratospheric air, ozone production from methane oxidation, the photooxidation of naturally emitted VOCs from vegetation, and the long-range transport of ozone formed from the photooxydation of anthropogenic VOCs and $NO_X$ emissions [19].

Present study is concerned with the monthly averages of total ozone over Arosa, Switzerland ($46.8^0$N/ $9.68^0$E) during the period from 1932 to 1971. The total ozone series of Arosa is the longest in the world. The measurements started in 1926 by F.W.P. Gotz, and since 1988, Metro Schweiz is responsible for operational measurements of total ozone at Arosa. Weiss (2000) [Ref 20] statistically analyzed the anthropogenic and dynamic contributions to total ozone over Arosa and developed regression models with North Atlantic Oscillation and Qusi-biennial Oscillation as predictors. Staehelin et al (1998a) [Ref 16] analyzed the total ozone series over Arosa with data homogenization technique. Bronnimann et al (2000) [Ref 2] discussed the variability of total ozone at Arosa considering the atmospheric circulation indices and established a close relation between total ozone time series and the atmospheric circulation at different height levels. Staehelin et al. (1998a, 1998b) [Refs 16,17] analyzed the trends of mean monthly-homogenized total ozone series over Arosa.

Present work is a bit deviated from the traditional statistical approaches and adopts Artificial Neural Network (ANN) as the predictive tool to forecast monthly mean total ozone on the basis of its own past values. Relevance of ANN in forecasting atmospheric and environmental pollution and the implementation procedure are described in the subsequent sections.

## 2. Artificial Neural Network with Backpropagation learning –an overview

Artificial Neural Networks (ANN) are biologically inspired network based on the organization of neurons and decision making process in the human brain [1]. In other wards, it is the mathematical analogue of the human nervous system. This can be used for prediction, pattern recognition and pattern classification purposes. It has been proved by several authors that ANN can be of great used when the associated system is so complex that the underline processes or relationship are not completely understandable or display chaotic properties. Development of ANN model for any system involves three important issues: (i) topology of the network, (ii) a proper training algorithm and (iii) activation function. Basically an ANN involves an input layer and an output layer connected through one or more hidden layers. The network learns by adjusting the inter connections between the layers. When the learning or training procedure is completed, a suitable output is produced at the output layer. The learning procedure may be supervised or unsupervised. In prediction problem supervised learning is adopted where a desired output is assigned to network before hand. The most popular learning algorithm used in prediction purposes is the back propagation algorithm (BPA) or the generalized delta rule. The BPA is a supervised learning algorithm that aims at reducing overall system error to a minimum [1,9]. This algorithm has made multilayer neural networks suitable for various prediction problems. In this learning procedure, an initial weight vectors $w_0$ is updated according to [10]:

$$w_i(k+1) = w_i(k) + \mu(T_i - O_i) f'(w_i x_i) x_i \qquad \ldots \qquad \ldots \qquad (1)$$

Where, $w_i \Rightarrow$ The weight matrix associated with i$^{th}$ neuron; $x_i \Rightarrow$ Input of the i$^{th}$ neuron; $O_i \Rightarrow$ Actual output of the i$^{th}$ neuron; $T_i \Rightarrow$ Target output of the i$^{th}$ neuron, and $\mu$ is the learning rate parameter.

The function $f(x)$ ,known as the activation function, in the RHS of the weight update equation is taken as the sigmoid function $f(x) = \dfrac{1}{1+e^{-x}}$ because of the form of its derivative. The graphical form of the function is given in Fig.04.

The ANN is learned with the weight update equation (1) to minimize the mean squared error given by [10]

$$E = \frac{1}{2}\left(T_i - O_i\right)^2 = \frac{1}{2}\left[T_i - f\left(w_i x_i\right)\right]^2 \qquad \ldots \qquad \ldots \qquad (2)$$

Since the purpose is to minimize the mean squared error, BPA algorithm is also called LMS method.

## 3. Artificial Neural Network in pollution study- a literature survey

Artificial Neural Network (ANN) has been widely used all over the world to predict time series pertaining to various kinds of pollution (e.g. Nunnari et al, 1998; Gardner and Dorling, 1998; Dorling et al, 2003) [Refs 5,6,13]. Comrie (1997) [Ref 3] compared the performance ANN over multiple regression models in predicting day-to-day ground-level ozone forecasting over a range of cities in USA using weather parameters as predictor and his ANN model gave somewhat better prediction than multiple regressions. Perez et al (2000) [Ref 14] implemented ANN in the form of Multilayer Perceptron to predict PM2.5 concentrations several hours in advance in Santiago, Chile. In that paper, the authors established superiority of non-linear ANN over linear model. Perez and Reyes (2001) [Ref 15] predicted particulate air pollution by developing ANN on the basis of information extracted from PM2.5 time series over Santiago, Chile. Kolehmainen et al (2001) [Ref 11] applied ANN in air quality forecasting by dealing with the periodic components. Corani (2005) [Ref 4] adopted ANN, pruned ANN, and Lazy Learning to predict ozone and PM10 over Milan, and reported that all the models are suitable for this prediction and in his research PM10 was predicted better than ozone. Gomez-Sanchies et al (2006) [Ref 7] developed an ANN model in the form of Multilayer Perceptron to forecast tropospheric ozone concentration over a site close to Valencia, Spain with vehicle emitted variables and climatological variables as predictors.

## 4. Neural Network in Total Ozone study –A review

Total ozone is a measure of the number of ozone molecules between the ground and the top of the atmosphere. In a more mathematical language, total ozone is simply the integral of the ozone concentration with respect to height.

The total ozone in mid latitude is a very complex phenomenon. As discussed earlier, total ozone is influenced severely by a number of meteorological factors, which are highly non-linear in nature. Total ozone includes both tropospheric and stratospheric ozone. Since the stratospheric ozone varies on different time scales, the influence on surface UV-B and related tropospheric ozone chemistry also acts on different time scales [21]. Thus, the total ozone time series is characterized by huge non-linearity attributed to meteorological variables, tropospheric ozone, and stratospheric ozone.

Thus, ANN is supposed to be a suitable device to study total ozone time series. Monge Sanz and Medrano Marques (2004) [Ref 12] developed ANN models for assessing long missing period of data of total ozone using North Atlantic Oscillation Index as predictor. Monge Sanz and Medrano Marques (2004) [Ref 12] attempted ANN in studying total ozone time series over Lisbon, Arosa, and Vigna di Valle all situated in mid-latitude. In this paper by Monge Sanz and Medrano Marques (2004) [Ref 12] reconstruction of time series by finding out the missing data points has been made using ANN in the form of Multilayer Perceptron Model using data between 1967 and 1980.

Present paper differs from Monge Sanz and Medrano Marques (2004) [Ref 12] and focuses on prediction of total ozone instead of reconstruction of time series using a much wider input window. Moreover, instead of using other predictor, this paper explores the past values of total ozone as predictor to forecast the future.

## 5. Data and analysis

Present paper deals with mean monthly total ozone concentration in Arosa, Switzerland between 1932 and 1971. The data are collected from http://www.robhyndman.info/TSDL/monthly/arosa.dat. The measurements are taken in Dobson Units (DU) (300 DU=1layer of 3mm if the whole ozone column is taken at the sea level with standard conditions)[8]. As the initial step towards data analysis the raw monthly data are schematically plotted. Some sample plots are presented in Fig.01 (a, b, c). The points of inflexion make it apparent that maximum concentrations of total ozone occur between February and May and a sharp increase in the concentration occurs from January to February. Average over the whole dataset is presented schematically in Fig.01 (d). This plot resembles almost the same pattern of the sample plots. Thus, it can be presumed that February to May is the peak period for total ozone concentration over Arosa.

In the next step we tried to examine any cyclic pattern in the dataset. To do the same autocorrelation coefficients [15] are computed up to a number of lags over the whole data set that is 40x12=480 months (data points). The autocorrelation function or the correlogram is presented in Fig.02. An oscillating pattern is apparent in the correlogram. The regular pattern proves that the time series of monthly mean of total ozone repeats a pattern in every 12 months cycle i.e. in a year's cycle. Since almost the same value of autocorrelation function comes out at lags separated by 12 lag points, it can be assume that the pattern of monthly mean total ozone time series bears more or less the same pattern in every year. Thus all the years under study can be considered on the same foot while framing the input matrix for the ANN.

In the next step, cross correlation coefficient [18] are computed between pairs of months. For example, there would be 40 February mean total ozone data and 40 March mean total ozone data would be there in the dataset and correlation would be computed between February and March data. There would be 12 such pairs in the dataset. All this correlations are computed and schematically presented in Fig. 03. It is found that the correlations are all positive. But their magnitudes are not as large to establish any linear association between pairs of data. Thus non-linearity in the dataset is recognized.

Since analysis of data shows that the period February-May is high in total ozone concentration and since high concentration is supposed to significantly influence the concentration of different trace gases in the troposphere [21], this period is chosen as the predictand.

## 6. Prediction using ANN

Prediction of total ozone in February-May using past mean monthly total ozone concentrations can be generalized as

y(Feb,Mar,Apr,May of year T)=*f*(xJ(T-1),xF(T-1),..,xN(T-1),xD(T-1),xJ(T)) …(3)

Where, in the RHS, xJ(T-1) to xD(T-1) imply monthly mean total ozone concentrations from January to December of year (T-1) and xJ(T) implies monthly mean total ozone concentration in year T. The LHS of equation (3) implies a set of four predictands.

Multilayer Perceptrons (MLP) with BPA are developed where information flows from input to output without feedback, and the connection weights are adjusted to minimize the MSE.

### 6.1 MLP with one hidden layer (Model 1)

As evident from equation (3), Model 1 contains 13 input variables. Since the study involves 39 years, there would be 38 rows in the input matrix. Thus, the input matrix is of order 38×13. Learning rate is fixed at 0.9 and momentum is set at 0.2. From the whole set, 75% i.e. 28 years are considered as training cases and 10 years are considered as the test cases. Only one hidden layer is taken. Since the number of adjustable parameters in a three-layered feed forward neural network with $n_i$ input units, $n_0$ output units, and $n_h$ hidden units is $[n_0+n_h(n_i+n_0+1)]$[14], and there are 28 training patterns, the number of hidden nodes can not be more than 2 nodes in the hidden layer. After a sequential learning using the set of equations presented in Section 2, the percentage of error of prediction (PE) is computed as [14]

$$PE = \frac{\langle |y_p - y_a| \rangle}{\langle y_{act} \rangle} \times 100 \qquad \dots \qquad \dots \qquad (4)$$

Where, $y_p$ is the predicted value, $y_a$ is the actual value, and < > means average over the test cases.

## 6.2 MLP with two hidden layers (Model 2)

Number of input variables, output variables, and the size of training and test sets are the same as the previous model. Since it is a two-hidden-layer model, there would be $[n_0+n_h(n_i+n_0+2)]$adjustable parameters and with 28 training patterns, there would be at the most 2 nodes in each hidden layer. In Model 2, each hidden layer is provided with 2 nodes. Learning rate is fixed at 0.9 and momentum is set at 0.2. After a sequential learning using the set of equations presented in Section 2, the percentage of error of prediction (PE) is computed as equation (4).

## 7. Results and discussion

Models 1 and 2 developed so far are now compared according to their goodness in predicting the spring-summer total ozone over Arosa. As far as the network topology is concerned, the models differ in the number of hidden layers only. But, the models have some significant differences in their predictive ability.

In both the models, sequential training has been executed up to 500 epochs. Both the models are found to converge with BPA learning. The pattern of the error surface for Model 2 is presented in Fig.05.

Actual and predicted total ozones in DU are presented in Fig 06 (a,b,c,d). Figures make it apparent that both of the model outputs agree very closely with the actual concentration of total ozone over Arosa. In the same figures, the individual errors in percentage are also presented. In all the cases, errors in prediction lie between 25% (both in training and test cases).

In case of February, percentage errors produced by Model 1 are greater than Model 2 in 72.5% cases. In March, Model 1 produced higher error than Model 2 in 32.5% cases and Model 2 produced higher error than Model 1 in 40% cases. In April, Model 1 produced higher error in 37.5% cases and Model 2 produced higher in 25% cases. In May, in 30% cases, Model 1 produced higher error and in 30% cases, Model 2 produced higher error. The results described in this paragraph correspond to both training and test cases.

Now, the test or the validation cases are considered only. The percentage of error of prediction (PE) and mean absolute error computed over the whole test set is presented in Fig.07 and Fig.08. These figures show that Model 2 performs significantly better than Model 1 in February, March, and April. But, in May, both of the models perform almost with same efficiency. Models differ most significantly in April. This implies that, increasing non-linearity in the model contributes very much to the prediction on total ozone in April. It may be inferred that April time series has the maximum degree of complexity. This is probably due to the fact that from April, total ozone over Arosa starts experiencing a falling trend.

## 8. Conclusion

From the study it is clear that time series pertaining to mean monthly total ozone over Arosa is a complex system and February to May is the time of maximum total ozone over this area. Artificial Neural Network in the form of Multilayer Perceptron with Backpropagation learning is found to be adroit for predicting the concentration of mean monthly total ozone over the station under study. Two neural net models are tested and both Single-hidden-layer and Two-hidden-layer ANN models with non-linear (Sigmoid) activation function are found suitable as predictive model. Between these two models, Two-hidden-layer ANN is found to be a better fit for predicting mean monthly total ozone over Arosa in the months of February, March, April, and May. Furthermore, April is identified as the month having the most complex pattern in the mean monthly total ozone over Arosa, Switzerland.

## References

[1] Acharya, U.R., Bhatt, P.S., Iyenger, S.S., Rao, A. and Dua, S., Classification of heart rate data using neural networks and fuzzy equivalence relation, *Pattern Recognition*, 36, 61-68 (2003).

[2] Bronnimann, S., Luterbacher, J. , Schmutz, C., Wanner, H., and Staehelin, J., Variability of total ozone at Arosa, Switzerland, since 1931 related to atmospheric circulation indices, *Geophysical Research Letters*, **27**, 2213-2216(2000).

[3] Comrie, A.C., Comparing Neural Networks and Regression Models for ozone forecasting, *J. Air & Waste Manage. Assoc*., **47**, 653-663 (1997).

[4] Corani, G., Air quality prediction in Milan: feed-forward neural networks, pruned neural networks and lazy learning, *Ecological Modelling*, 185, 513-529 (2005)

[5] Dorling, S., Foxall, R., Mandic, D. and Cawley, G., Maximum likelihood cost functions for neural network models of air quality data. *Atmospheric Environment*, **37**, 3435-3443 (2003)

[6] Gardner, M. and Dorling, S., Artificial neural network (the multilayer perceptron) - a review of applications in the atmospheric sciences, *Atmospheric Environment*, **6**, 2627-2636 (1998)

[7] Gomez-Sanchis, J., Martin-Guerrero, J.D., Soria-Olivas, E., Vila-Frances, J., Carrasco , J.L. and del Valle-Tascon, S., Neural networks for analysing the relevance of input variables in the prediction of tropospheric ozone concentration, Preprint submitted to Elsivere, available in http://www.uv.es/jdmg/ozone_atm_env.pdf (2006).

[8] Institute for Atmospheric and Climate Science (IACS), Total ozone series in Arosa,,Switzerland, http://www.iac.ethz.ch/en/research/chemie/tpeter/totozon.html#results (2004)

[9] Kamarthi, S.V. and Pittner, S, Accelerating neural network training using weight extrapolation, *Neural Network*, **12**, 1285-1299(1999).

[10] Kartalopoulos, S.V., *Understanding Neural Networks and Fuzzy Logic- Basic Concepts and Applications*, (Prentice Hall, New-Delhi, 2000).

[11] Kolehmainen, M., Martikainen, H., and Ruuskanen, J., Neural networks and periodic components used in air quality forecasting, *Atmospheric Environment*, **35**, 815-825 (2001)

[12] Monge Sanz, B. M. and Medrano Marques, N. J., Total ozone time series analysis: a neural network model approach, *Nonlinear Processes in Geophysics*, **11**, 683-689 (2004)

[13] Nunnari, G., Nucifora, M. and Randieri, C., The application of neural techniques to the modelling of time-series of atmospheric pollution data, *Ecological Modelling* , **111**, 187-205 (1998)

[14] Perez, P., Trier, A. and Reyes, J., Prediction of $PM_{2.5}$ concentrations several hours in advance using neural networks in Santiago, Chile, *Atmospheric Environment*, **34**, 1189-1196 (2000)

[15] Perez, P. and Reyes, J., Prediction of particulate air pollution using neural techniques", *Neural Computing and Application*, **10**,165-171(2001)

[16] Staehelin, J., Renaud, A., Bader, J., McPeters, R., Viatte, P., Hoeg-ger, B., Bugnion, V, Giroud, M., and Schill, H., Total ozone series at Arosa (Switzerland): Homogeneization and data comparison. *J. Geophys. Res*., **103, D5**, 5827-5841 (1998a).

[17] Staehelin, J., Kegel, R., and Harris, N. R. P., Trend analysis of the homogenized total ozone series of Arosa (Switzerland), J. *Geophys. Res*.**,** **103***,* 8389-8399 (1998b).

[18] Wilks, D.S., *Statistical Methods in Atmospheric Sciences* (Academic Press, USA, 1995)

[19] Wolff, G., Air Pollution, In: *Encyclopedia on Environmental Analysis and remediation*, Edited by R. A. Meyer (Wiley, 1998).

[20] Weiss, A.K., *Anthropogenic and Dynamic Contributions to Ozone Trends of the Swiss Total Ozone Umkher and Balloon Sounding Series*, Ph.D. Work (GCA-Verlag, 2000).

[21] Zanis, P., Kourtidis, K., Balis, O. and Zerefos, C., On the Photochemical Relation between Surface and Total Ozone ? A Case Study during the PAUR II experiment at Crete, Greece, Available in

http://ies.jrc.cec.eu.int/Units/cc/events/torino2001/torinocd/Documents/Anthropogenic/AP81.htm

(2001).

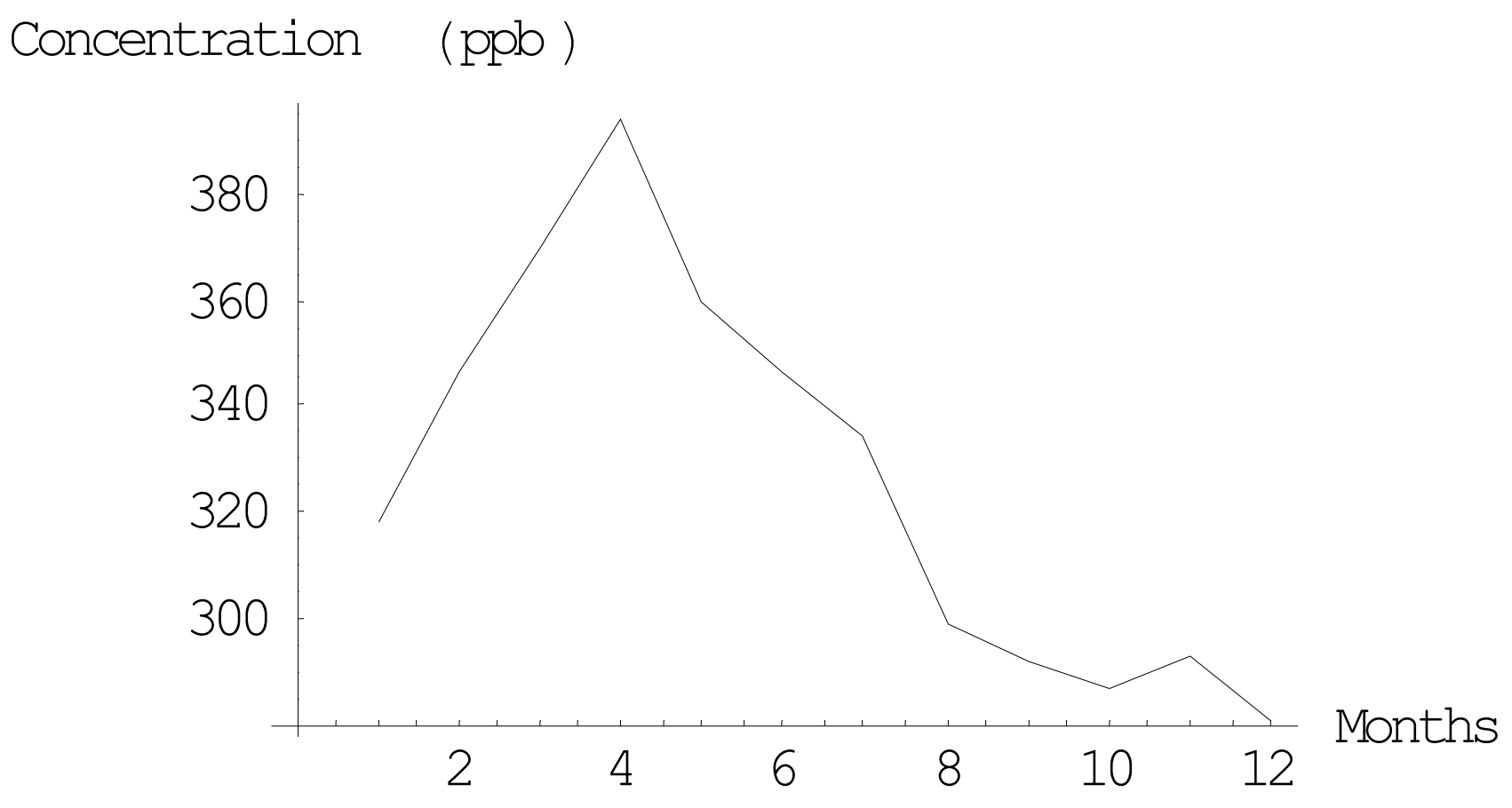

Fig01a - Monthly concentration of Ozone in 1932.

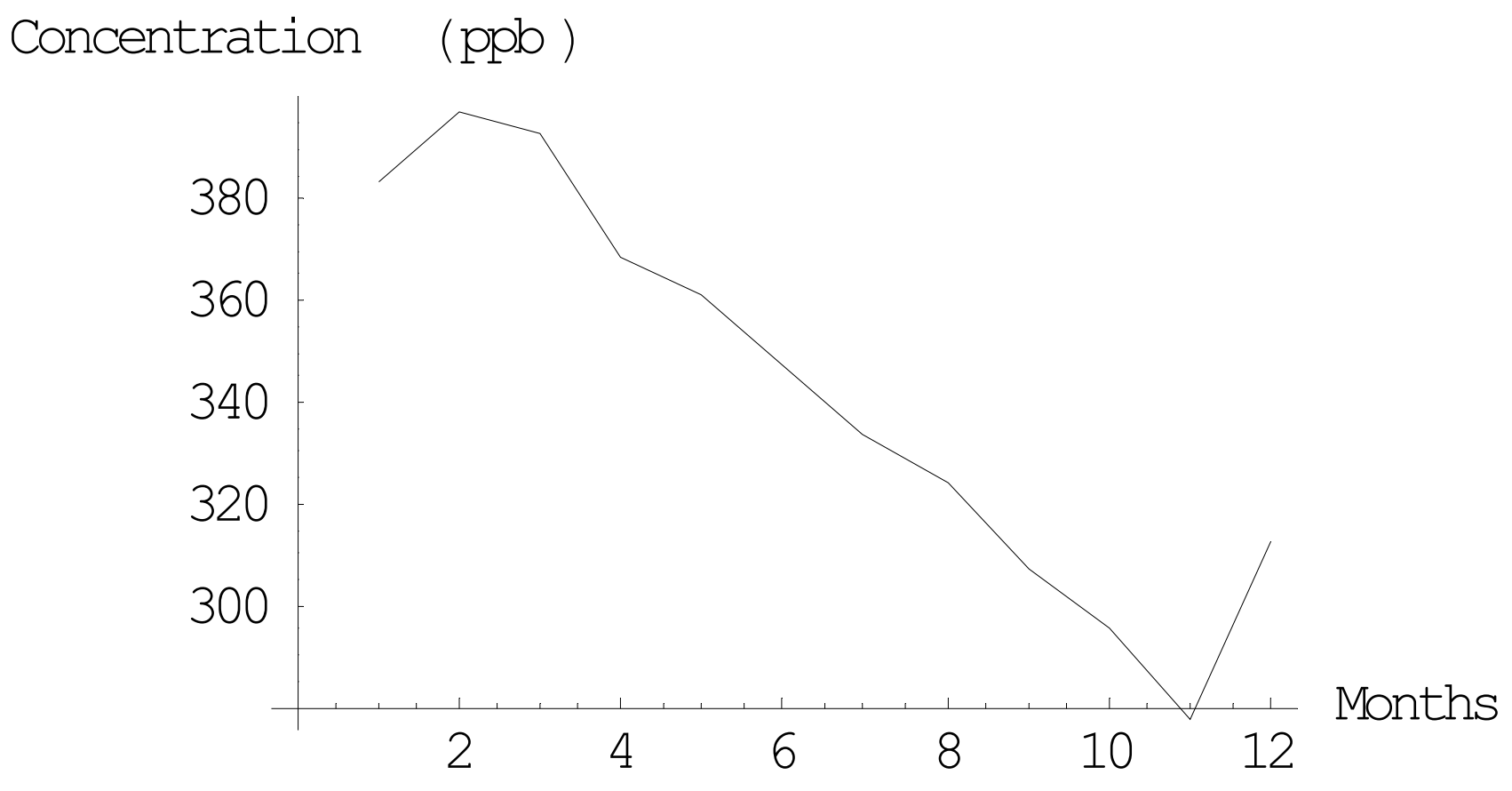

Fig01b - Monthly concentration of Ozone in 1947.

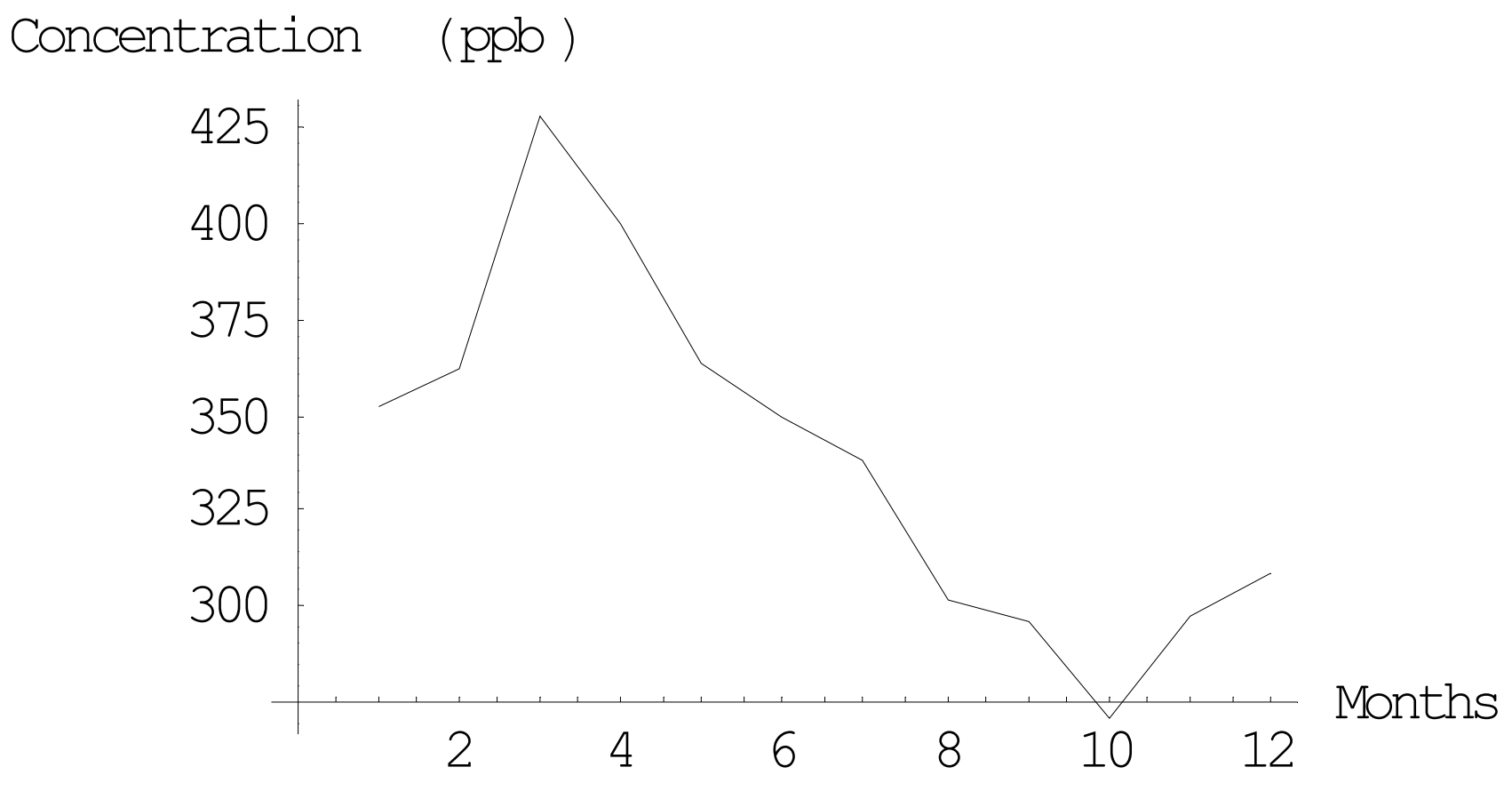

Fig01C – Monthly concentration of Ozone in 1962.

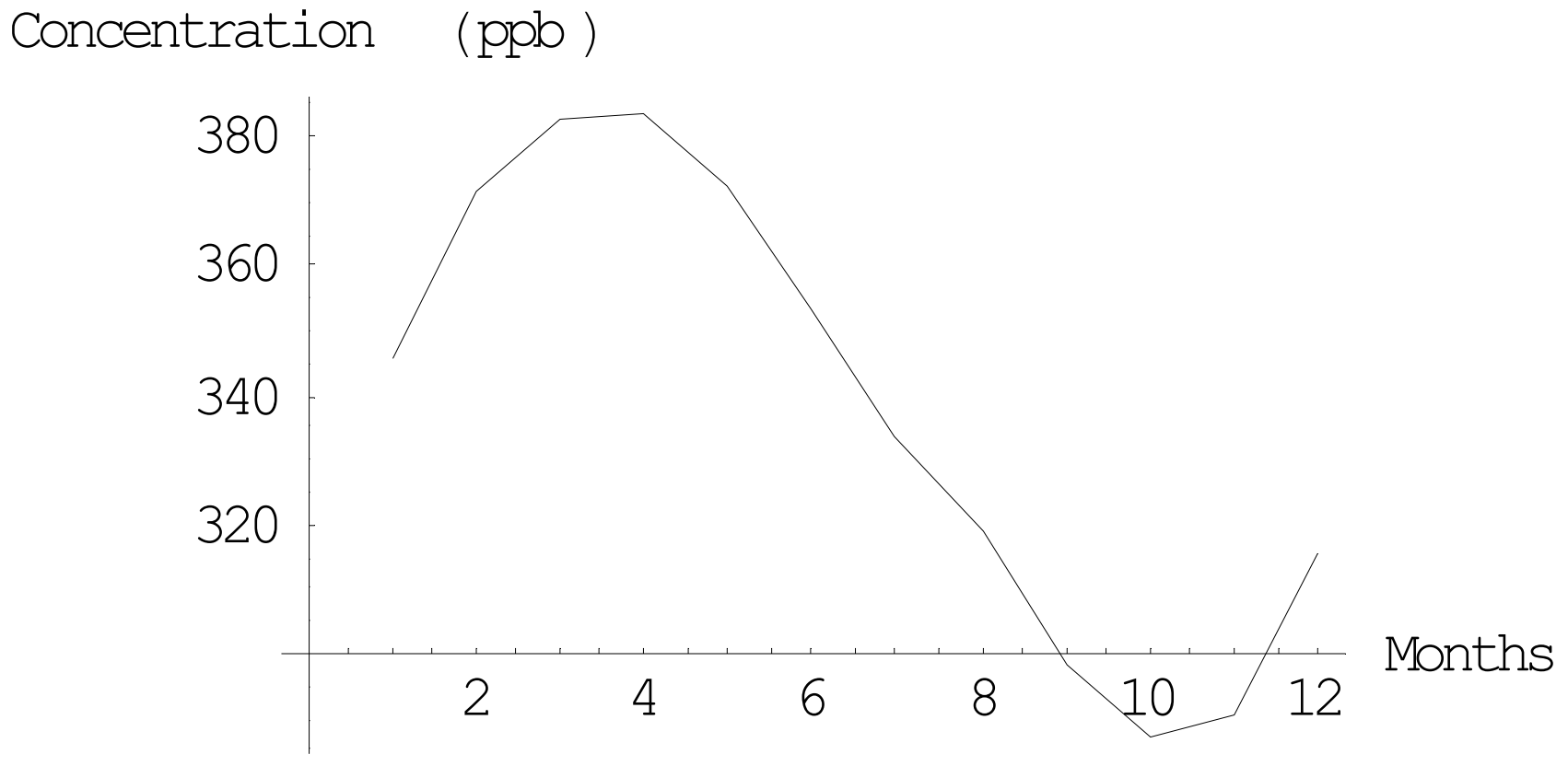

Fig01d – Monthly concentration of Ozone averaged over 1932 – 1971

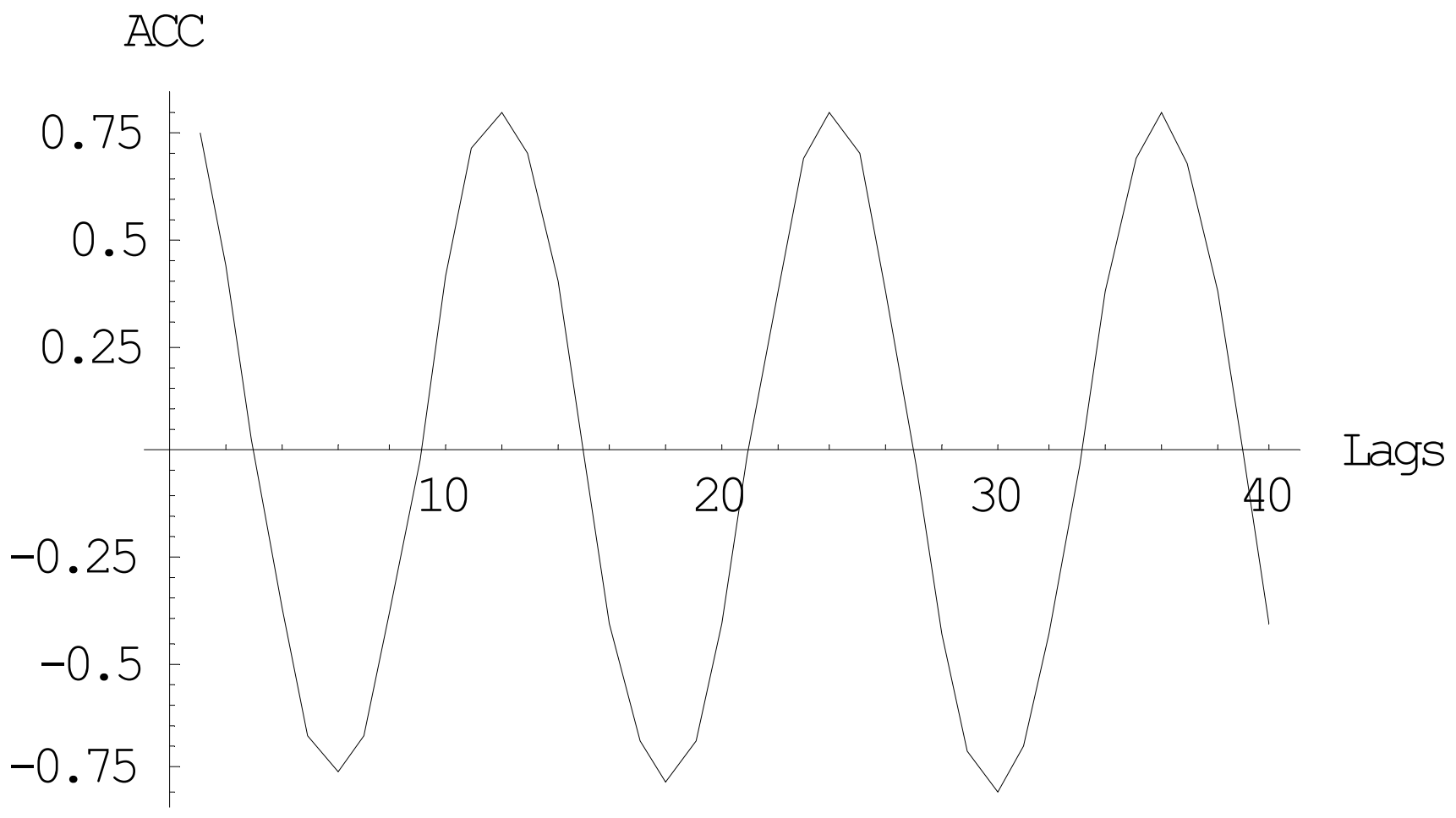

Fig.02 – Autocorrelation function for the monthly Ozone concentrations over Arosa between 1932 and 1971.

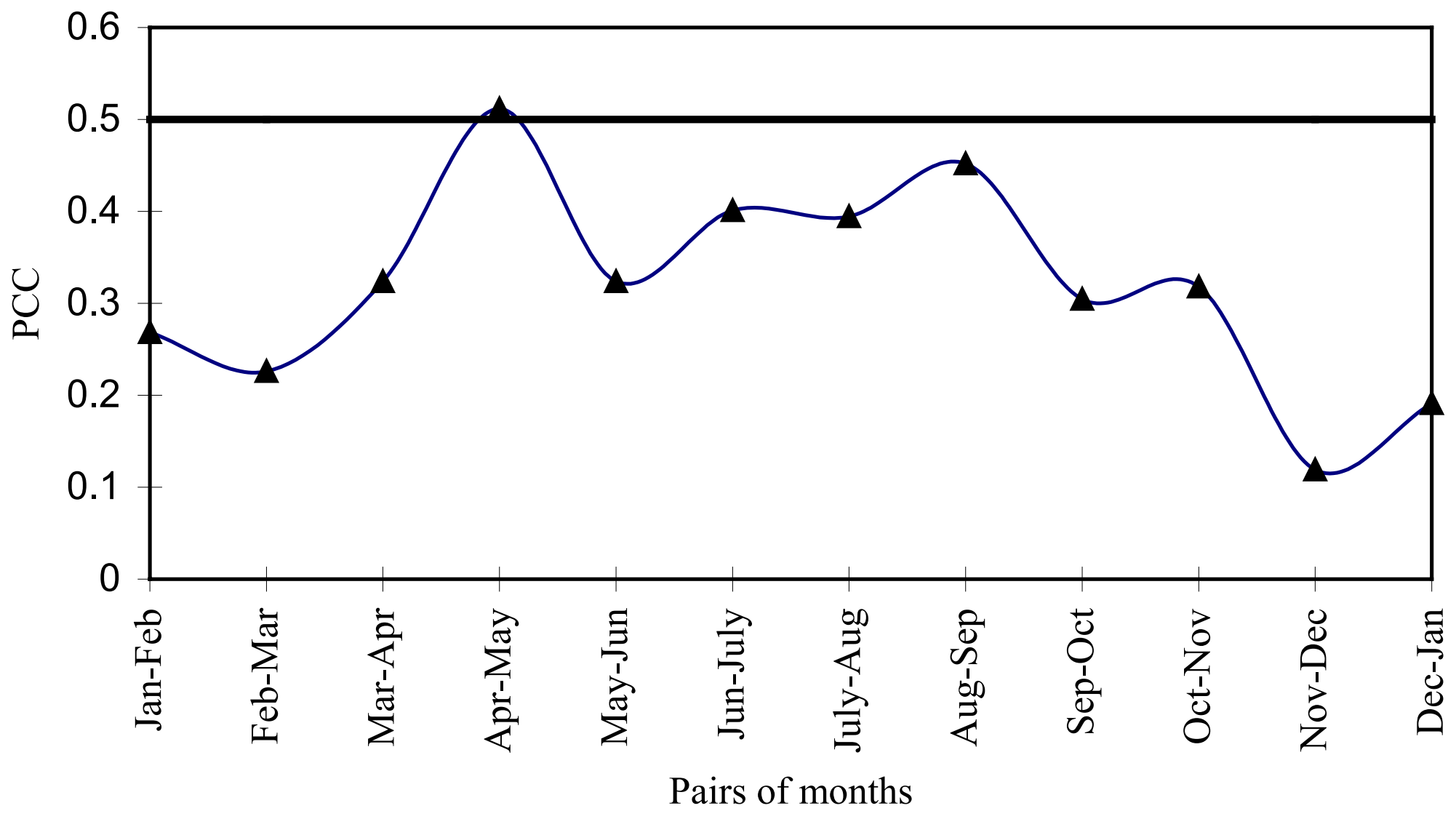

Fig.03-Pearson correlation coefficient (PCC) between monthly mean total ozone over Arosa corresponding to pairs of months computed over 40 years.

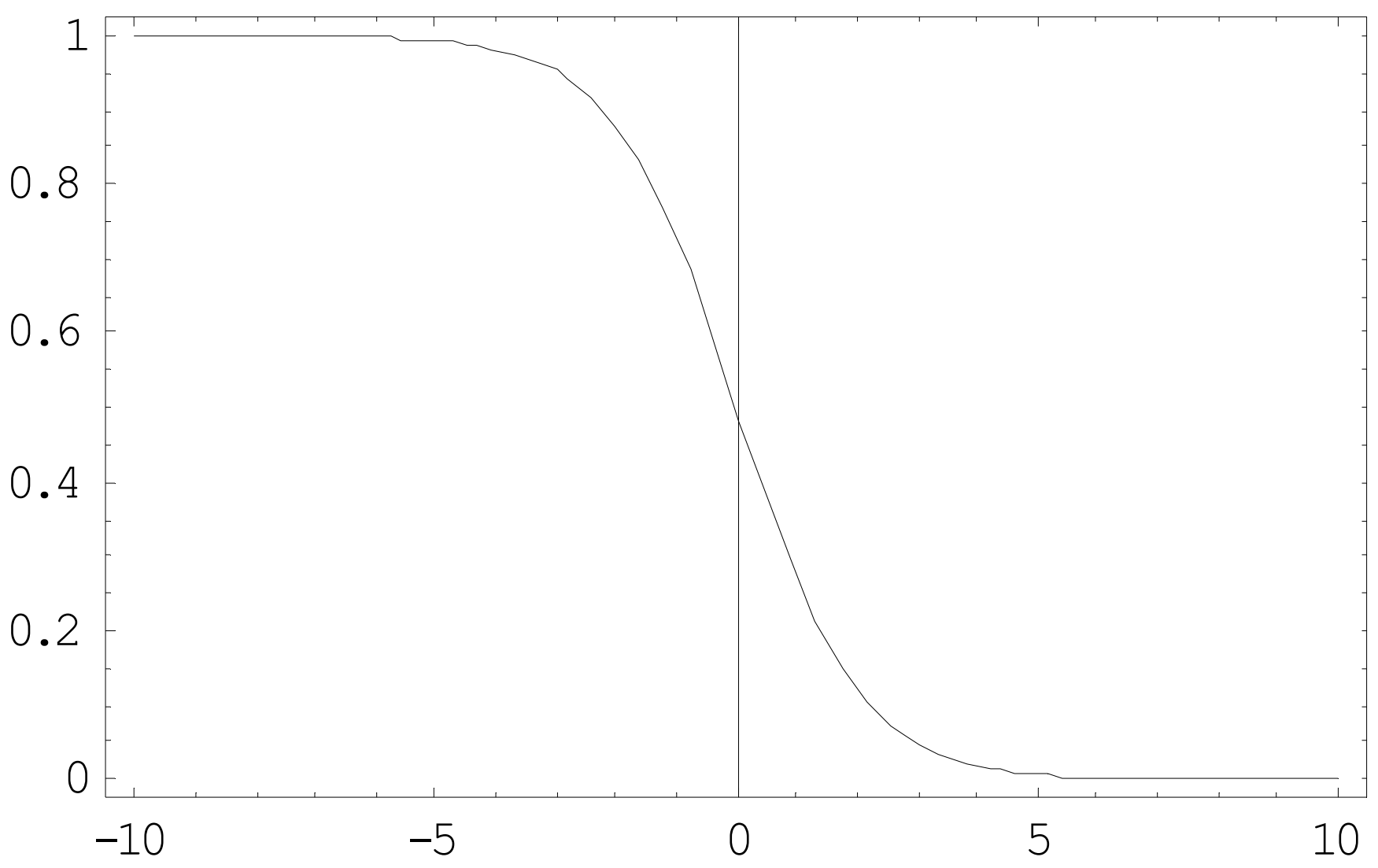

Fig.04- The sigmoid function

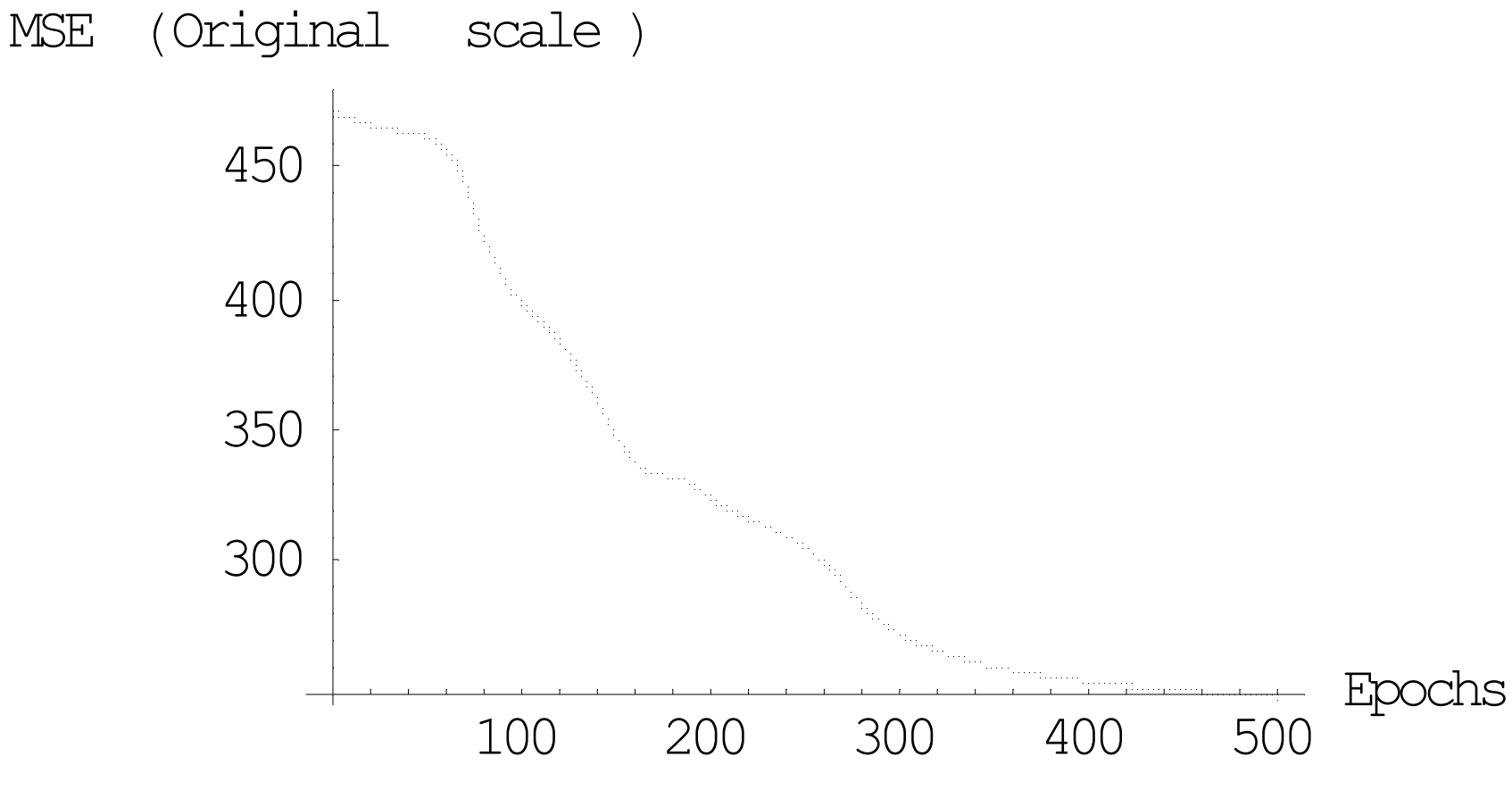

Fig05 – Change in the mean squared error (MSE) with number of epochs in the training phase of the FFNN with two hidden layers and four output variables (Model2).

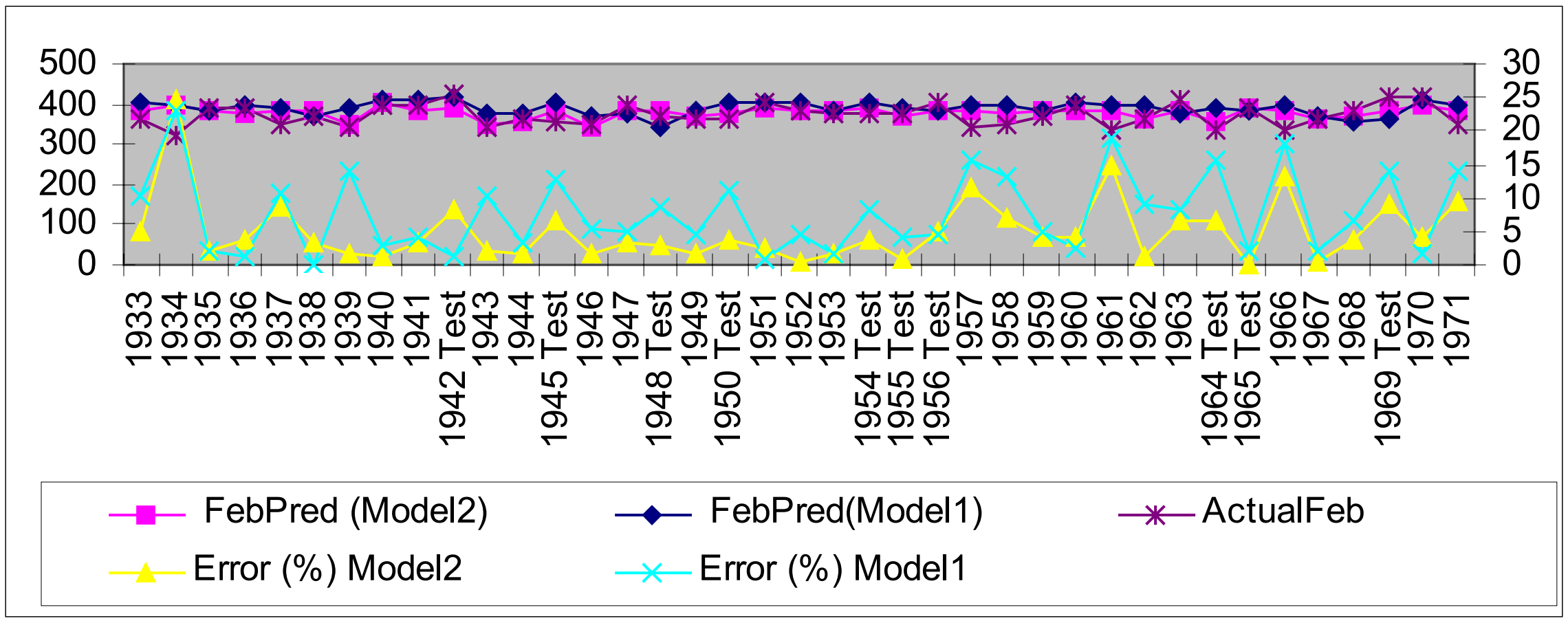

Fig.06 (a) – Actual total ozone and prediction from ANN models in February with percentage errors of prediction. Training and test cases are presented.

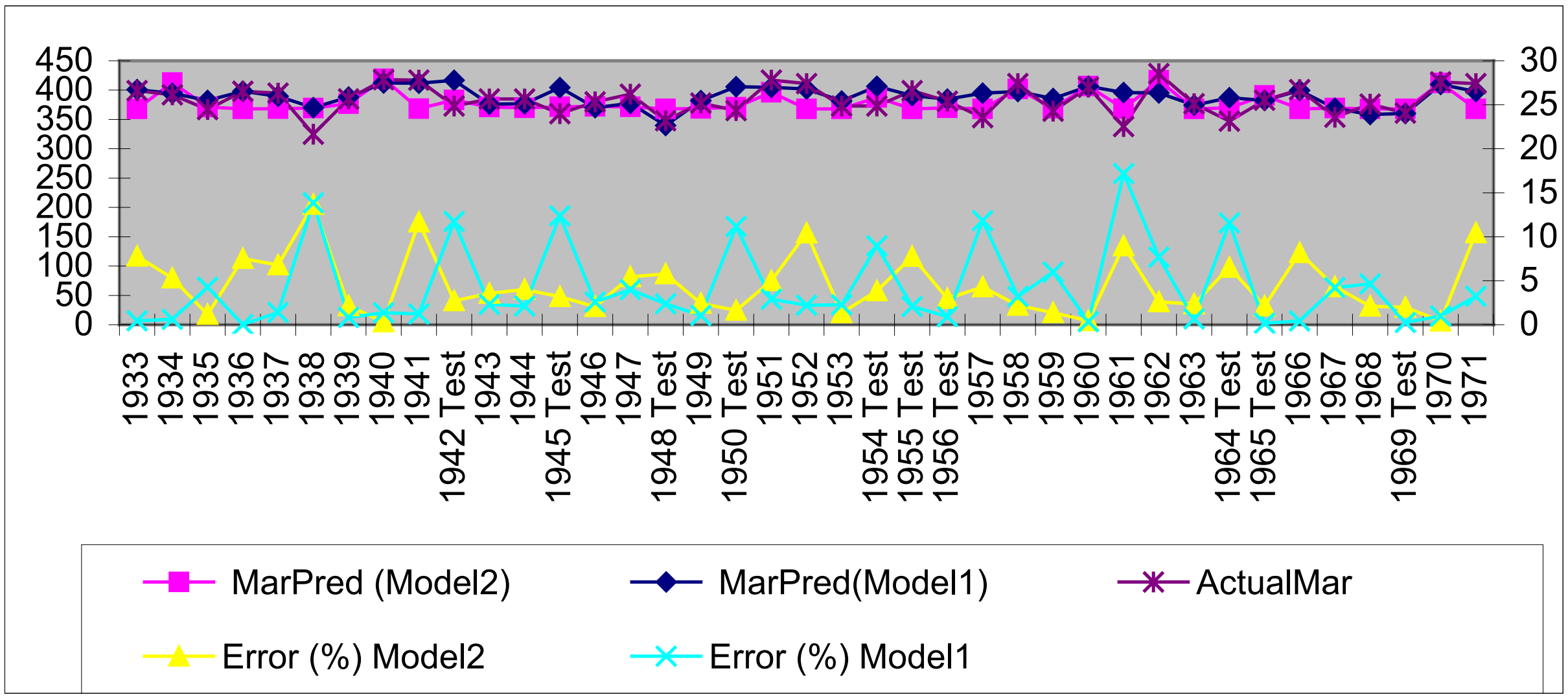

Fig.06 (b) – Actual total ozone and prediction from ANN models in March with percentage errors of prediction. Training and test cases are presented.

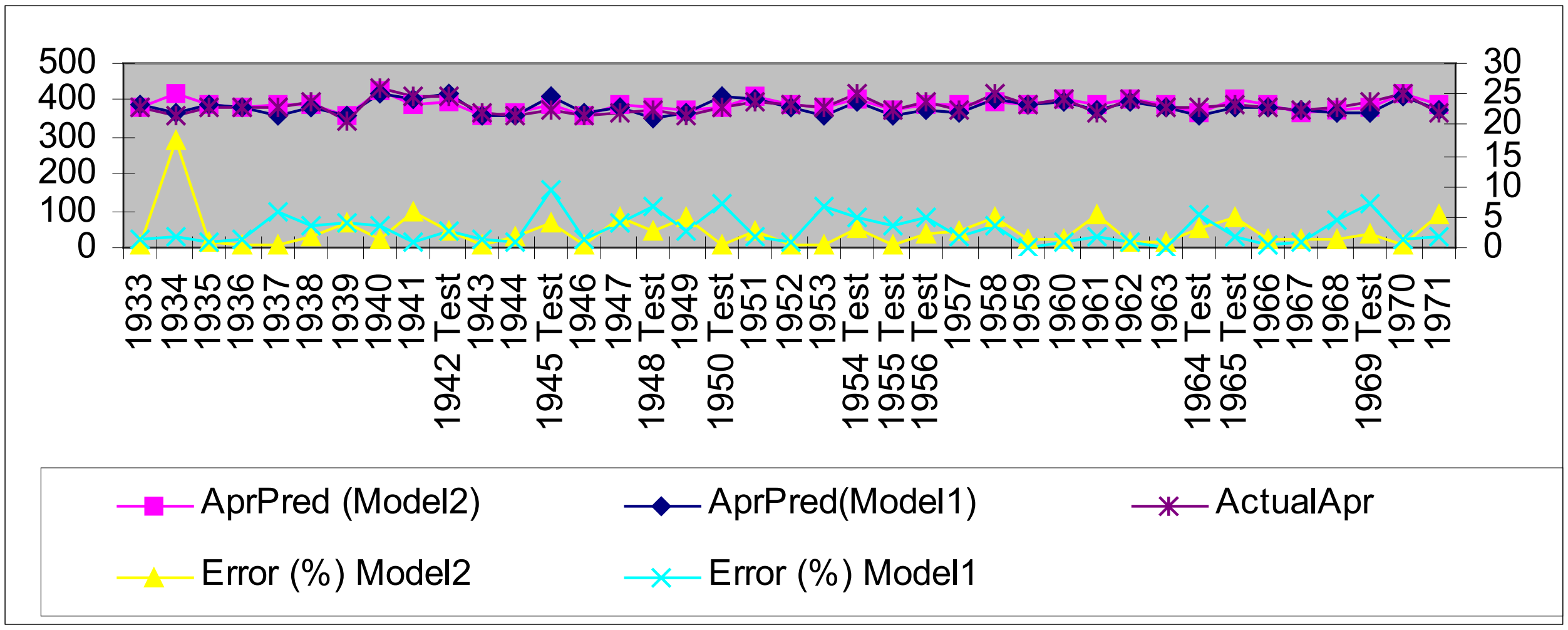

Fig.06 (c) – Actual total ozone and prediction from ANN models in April with percentage errors of prediction. Training and test cases are presented.

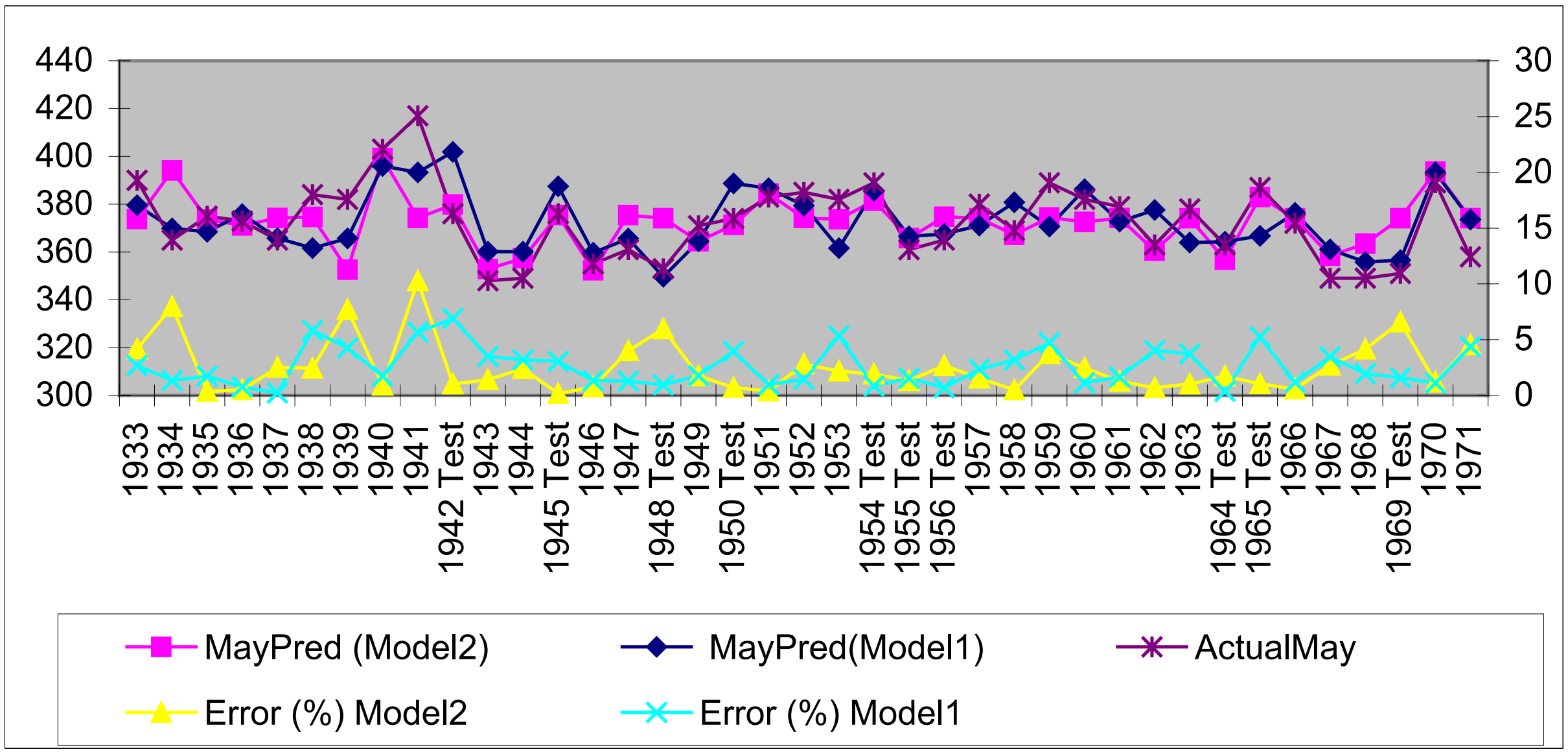

Fig.06 (d) – Actual total ozone and prediction from ANN models in May with percentage errors of prediction. Training and test cases are presented.

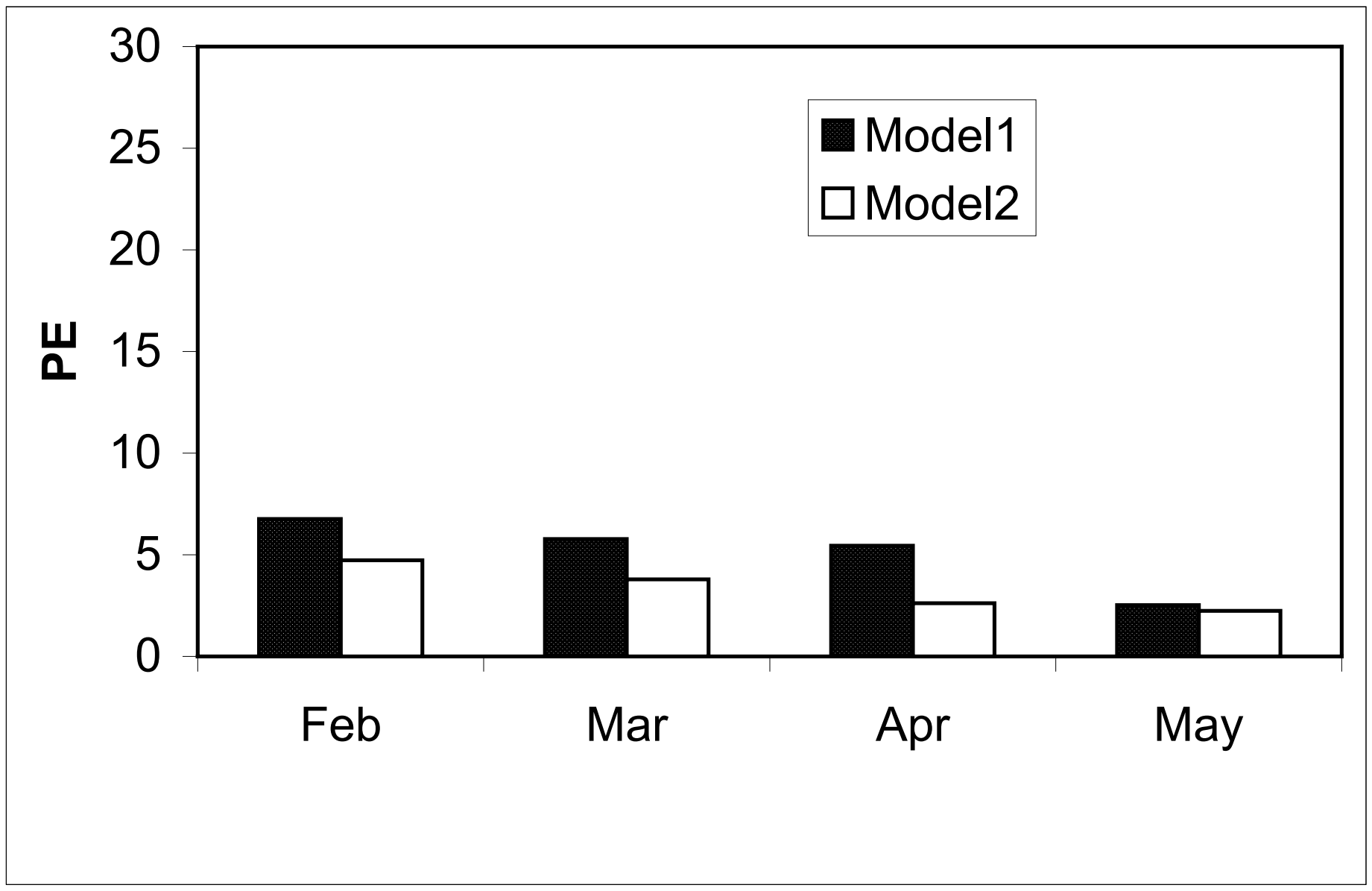

Fig.07- Schematic of the PE from model1 and model2 (Test cases)

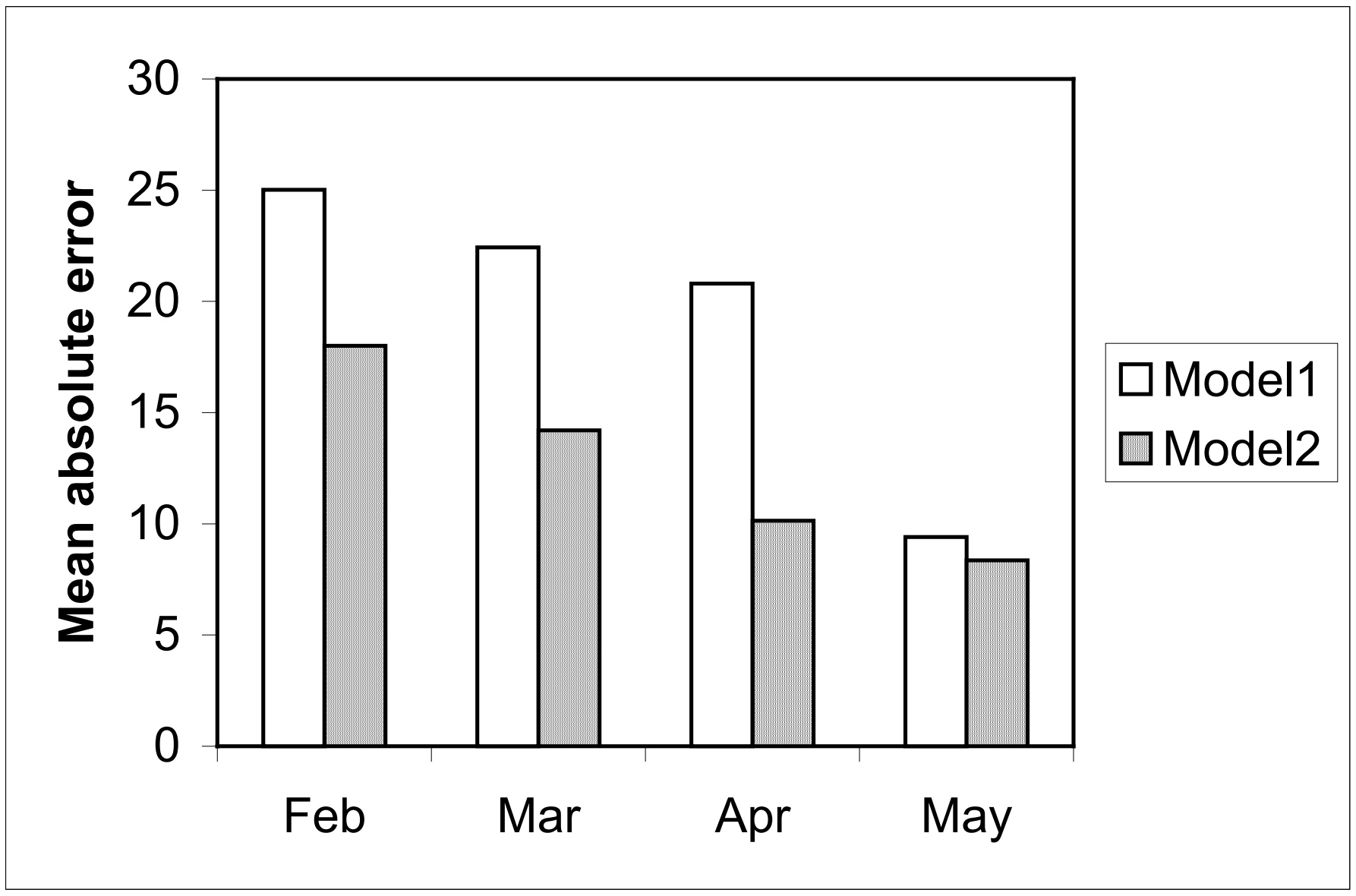

Fig.08- Schematic of the mean absolute error of prediction from model1 and model2 (Test cases)